\documentclass[square]{ws-procs975x65}

\begin{document}

\title{What can the Higgs tell us about UV physics?\footnote{\it Talk given at XXIX-th International Workshop on High Energy Physics at IHEP, Protvino.}}

\author{A. K. KNOCHEL}

\address{Institut f\"ur Theoretische Teilchenphysik und Kosmologie, RWTH Aachen,\\
Aachen, Germany\\
E-mail: knochel@physik.rwth-aachen.de\\
www.rwth-aachen.de}

\begin{abstract}
After the discovery of a Higgs-like boson with a mass of $m_h\approx 125$ GeV 
at the LHC, we can now attempt
to draw conclusions about physics beyond the Standard Model. I argue that 
there are several hints towards new physics at intermediate scales $\Lambda
\gtrsim 10^8$ GeV. I review a class of stringy models with intermediate
scale SUSY which relate the observed Higgs mass to symmetries 
of the Higgs sector. I then discuss radiative corrections to $m_h$, unification, dark matter
and the possibility of classically unstable UV completions in these models. 
\end{abstract}


\bodymatter

\section{Introduction}\label{aba:sec1}
The properties of the new particle\cite{Chatrchyan:2012ufa,Aad:2012tfa} at $m\approx 125$ GeV discovered by the 
ATLAS and CMS experiments at the CERN LHC match within experimental
uncertainties those predicted by the minimal Higgs
sector of the Standard Model\cite{Falkowski:2013dza,Plehn:2012iz,Djouadi:2013qya,Corbett:2012ja}. These
experimental uncertainties are already small enough to be a nontrivial
test of the SM Higgs hypothesis. Alternative spin assignments as well as parity
assignments are experimentally disfavored, and those couplings of the new boson
to SM states which have already been measured agree with predictions closely enough to 
warrant the name ``Higgs boson''. We therefore have an indirect (albeit
model-dependent) measurement of the last unknown parameter of the minimal
Standard Model - the quartic Higgs coupling.  
For the sake of this talk, I will assume that
the new boson is indeed the SM Higgs in the sense that deviations from the SM Higgs
sector are suppressed by a large new physics scale significantly above a TeV.
\begin{figure}
\begin{center}
\includegraphics[width=6cm]{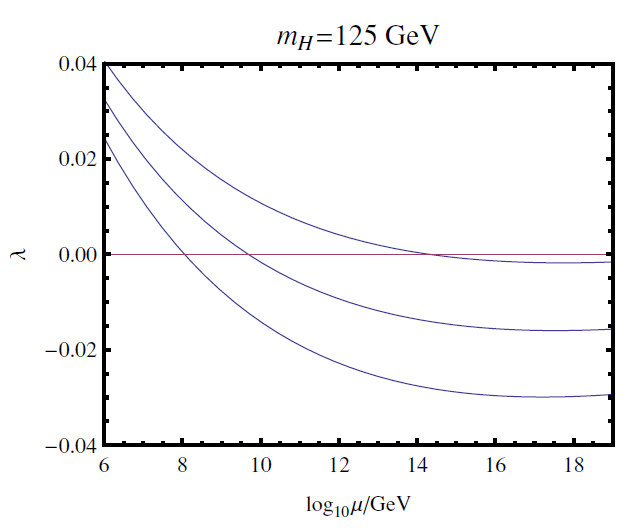}
\end{center}
\caption{The 2-loop renormalization group flow of the Higgs quartic coupling in the 
Standard Model for values of the top quark mass of $m_t=170.7, 172.9, 175$ from upper to lower line. The quartic
coupling vanishes at or near the Planck scale only for very low values of the top quark mass. Around
the current PDG central value, the sign change takes place at intermediate scales. }
\end{figure}

In absence of evidence for
other new physics at the LHC8 and other colliders, and only indirect or
unspecific experimental evidence from other observations, one can ask how strongly
the physics at the TeV scale really deviates from the electroweak Standard
Model. What is the scale at which a radical departure from the minimal SM is to be
expected, and of what type is this departure? Since before the definite discovery of the
new boson, it has been noted by several authors (see for example
\cite{EliasMiro:2011aa,Holthausen:2011aa,Degrassi:2012ry,Cabrera:2011bi }) that
the top quark Yukawa coupling and the Higgs quartic 
coupling, interpreted within the SM, take very peculiar values in nature: we live
very close to a ``critical line'' in the $m_h-m_t$ plane which separates the 
parameter region of absolute vacuum stability (high Higgs masses and low top masses) 
from a region of instability in which the SM predicts our vacuum $\langle h\rangle \sim 175$ GeV 
to be metastable or even unstable at cosmological time scales. Since this 
analysis is predicated on taking the SM Higgs sector at face value up to 
very high energy scales with only mild modifications, and since we still 
appear to be in the region of sufficient stability on cosmological time scales, no
absolutely imperative conclusions can be drawn from it. However, in light of
absence of evidence for other new physics at the LHC,
nature's location in the $m_h-m_t$ plane could be taken as a hint that
the SM Higgs sector might remain essentially unmodified up to scales far beyond a TeV.
If significant modifications of the minimal Higgs sector such as scalar singlet
extensions, 2HDM (SUSY or not), or
compositeness exist at the TeV scale, the stability diagram is meaningless and
our position on the ``would-be'' critical line a mere coincidence. 

When interpreted
within the intermediate scale SUSY scenario which we propose \cite{Hebecker:2012qp,Hebecker:2013lha,Ibanez:2012zg,Ibanez:2013gf}, the
vanishing of the quartic coupling at this mass scale can be explained from stringy symmetries
of the Higgs sector, and a 
connection of the observed Higgs mass is established to other phenomena in nature 
which point towards intermediate scales of new physics such as neutrino masses (and possibly
leptogenesis),
axion dark matter and gauge unification. The preferred axion decay constant for dark matter 
is around $f_a\approx 10^{12}$ GeV, where higher values can be accommodated if the initial
misalignment angle is $\theta\ll 1$, and smaller ones if there are other DM sources. These models are
currently being tested, e.g. by the ADMX experiment. Gauge unification
can be easily achieved in our scenario without low scale SUSY, for example in the presence of GUT breaking 
gauge flux in type IIB compactifications. In Fig. 2 of \cite{Ibanez:2012zg} this
coincidence of scales is nicely illustrated for a specific model.

I argue that after LHC8, there should be renewed efforts to think 
about UV completions at {\it intermediate} scales and what they might entail for 
(stringy) model building, cosmology, HEP and dark matter experiments. 

\section{Shift symmetric Higgs sectors}
Recently, we have proposed \cite{Hebecker:2012qp,Hebecker:2013lha} how the projected vanishing of
the quartic coupling at intermediate scales may be connected to stringy UV completions. They predict a vanishing tree level Higgs 
quartic coupling at the soft breaking scale due to an approximate shift symmetry\footnote{It was since proposed \cite{Ibanez:2012zg} to 
realize this situation using an approximate $\mathbb Z_2$ parity rather
than shift symmetries.} 
\begin{equation}
H_u\longrightarrow H_u+c,\qquad \overline H_d \longrightarrow \overline H_d - c
\end{equation}
in the Higgs sector. It restricts the leading order lowest
dimension K\"ahler potential to be of the form
\begin{equation}
\mathcal K \sim f(X) |H_u+\overline H_d|^2
\end{equation}
where $f(X)$ encodes the moduli dependence of the K\"ahler
function. An immediate consequence of this is that $\tan\beta=1$ and the
SM Higgs doublet lies along a flat direction of the electroweak D-term
\begin{equation}
V\sim \frac{g_1^2+g_2^2}{8} (|H_u|^2-|H_d|^2)^2+\dots\,.
\end{equation}
There is therefore no quartic self coupling at tree level.

Such shift symmetries have been known for quite some time
to appear in Heterotic orbifold compactifications \cite{LopesCardoso:1994is,Antoniadis:1994hg,Brignole:1995fb,Brignole:1996xb} and simpler field theoretic
models \cite{Choi:2003kq,Brummer:2009ug}, where they are essentially a remnant of higher-dimensional gauge
invariance\footnote{However, one has to be careful since it depends
on the details of the compactification whether a shift symmetry is
actually realized in terms of the variables of the 4D K\"ahler potential \cite{Hebecker:2013lha}. 
For example, both components of the complex Wilson line moduli on D7 branes transform
nonlinearly under certain gauge transformations, which would naively entail
that they drop out of the K\"ahler potential entirely if it were shift-symmetric with
respect to both.}.
In \cite{Hebecker:2013lha} we argue that in type II models, analogous situations 
can arise not only for type IIA Wilson line Higgs sectors but also for a type IIB bulk
Higgs on D7 branes. In the remainder of this talk I want to concentrate on certain field-theoretic aspects of
these models, and in particular on the effective field theories below the
compactification scale.
\section{Radiative Corrections to the Weak Scale and the Higgs Mass}
The hierarchy problem is not obviously present in 
the SM in regularization/renormalization schemes such as $\overline{MS}$/$\overline{DR}$ or functional
renormalization group techniques
which avoid the introduction of a hard scale-invariance breaking
cutoff\cite{Bardeen:1995kv,Shaposhnikov:2009pv}. It reappears when
new particles coupling to the SM exist far beyond the TeV scale. It is
conceivable that
the hierarchy problem is somehow remedied at this high scale (in contrast to SUSY which must be present
far below the high scale in order to work as a remedy), but no mechanism is
presently known to us.  The relevant scale in the shift symmetric SUSY
scenarios is $\Lambda_{\lambda=0}/4 \pi\ll M_{Pl}$, giving us a fine tuning
measure which is large but nevertheless up to $\sim 23$ orders of magnitude less severe than the naive 
cutoff-based estimate in the SM, $M_{pl}^2/m_W^2$. 
Since we have a theory prediction for the quartic coupling, once the electroweak
scale is set to the measured value, the Higgs mass is fixed as well. We now
want to consider the radiative corrections to this ratio $m_h/m_W$, i.e. to the physical Higgs mass.

\begin{figure}
\begin{center}
\begin{picture}(500,160)
\put(0,50){\rotatebox{90}{$m_h/GeV$}}
\put(140,0){$\log_{10}(m_S/GeV)$}
\put(15,10){
\includegraphics[width=5cm]{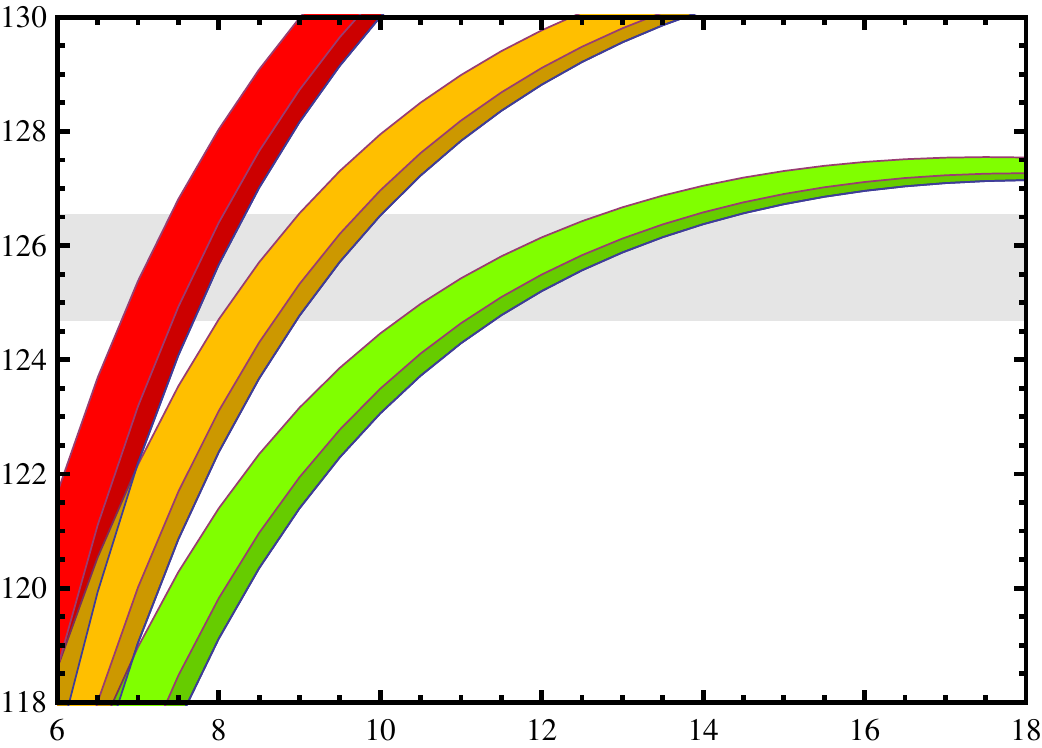}\qquad
\includegraphics[width=5cm]{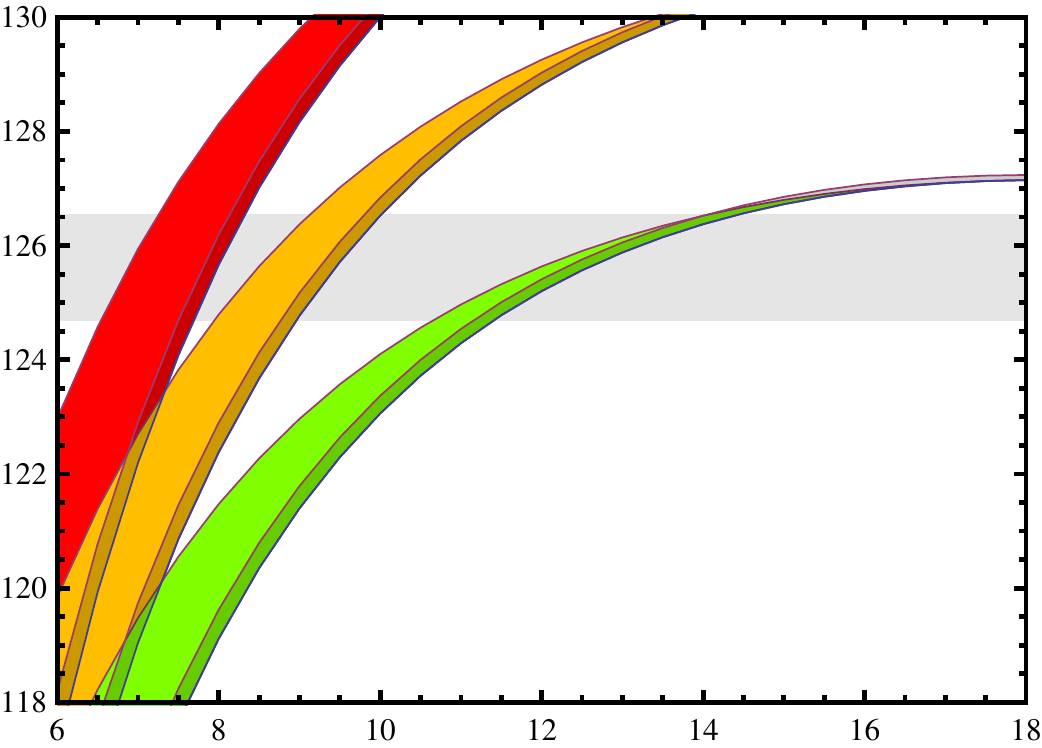}}
\end{picture}
\end{center}
\caption{The impact of squark decoupling corrections to the quartic Higgs
coupling (left) and shift/exchange symmetry violation (right) on the physical
Higgs mass.  The narrow dark(broad light) bands are for $X_t^2=m_S^2\,(6
m_S^2)$ for the decoupling contributions from top partners, and $m_C=10^2\, m_S
(\sqrt{m_S m_{Pl}})$ for the shift symmetry violation. The top quark masses are
$m_t=175.5,173.5,171.5$ from upper (red) to lower (green) band. The scale $m_S$
should be understood as the effective SUSY scale.\label{fig_thresholdhiggsmasses}}
\end{figure}

The SM effective potential for the Higgs and consequently the relation 
between the $\overline{MS}$ quartic coupling and $m_h$ as well as the running
of the quartic coupling are well known to NNLO. 
We are now concerned with 
the corrections to the quartic coupling at the high scale of new physics. 
There are two main contributions: i) corrections to $\tan\beta=1$ and thus 
to the tree level quartic coupling; ii) radiative corrections to the quartic 
coupling itself. The former is suppressed by one additional loop factor, but
it can be log-enhanced by large hierarchies between the soft scale and the
string compactification scale, and therefore competes with the 1-Loop radiative
corrections. \\
{\bf i) Corrections to $\tan\beta=1$} or equivalently to $\cos 2\beta=0$ arise when the ``shift-symmetric'' Higgs mass 
matrix $m_1^2 = m_2^2 = B\mu$ with an exactly massless eigenstate receives
radiative corrections which destroy this degeneracy. This is generally the case
if the top mass comes from $\mathcal W\sim H_u Q T$ at the renormalizable level.
One can control this radiative violation of shift symmetry
by dialling the soft breaking parameters in order to obtain an
$\mathcal O(100)$ GeV eigenvalue. However, the resulting Higgs mass matrix will
generally yield $\cos2\beta=\epsilon$, where $\epsilon\ll 1$ depends on the
details of our parameter choice. We can give a good estimate of its
magnitude. The resulting tree level quartic coupling is \cite{Hebecker:2012qp}
\begin{equation}\delta \lambda_{SV}(m_S)\sim C\,\frac{g_2^2+g_1^2}{8}\,\Big|\frac{6 \overline{y_t^2}}{16 \pi^2}\log\left(\frac{m_S}{m_C}\right)\Big|^2\,.
\label{shiftdeltalambda}
\end{equation}
where $C$ is an $\mathcal{O}(1)$ constant, and $m_S$ and $m_C$ are the soft 
breaking and compactification scales.\\
{\bf ii) Corrections to $\lambda$} at the high scale arise from loop diagrams with four external Higgs fields 
and heavy internal lines. We operate in the decoupling limit of the MSSM, where the masses of the
extended Higgs sector and the superpartners are spread around the soft scale. In the limit $\cos 2\beta\ll 1$,
the resulting corrections to $\lambda$ are given by\cite{Hebecker:2013lha}
\begin{eqnarray}
\delta \lambda &=& \frac{3 y_t^4}{16 \pi^2}\Big[
\frac{X_t^2}{m_{\tilde t}^2}\Big(1-\frac{X_t^2}{12 m_{\tilde
t}^2}\Big)+2\log(\frac{m_{\tilde t}}{m_S})\Big] 
-\frac{1}{16 \pi^2} \frac14 \tilde b_\lambda \log\frac{m_A}{m_S} \nonumber \\
&+&\frac{\tilde b_\lambda}{16 \pi^2}\,\left [\log \frac{\mu}{m_S} + \frac{(r-1)(r+1)^2 +2(r-3) r^2 \log r}{2 (r-1)^3}\right]\,
\end{eqnarray}
where 
$\tilde b_\lambda= \frac12(-g_1^4-2 g_1^2 g_2^2 -3 g_2^4),\, \frac{M_1}{\mu}= \frac{M_2}{\mu}=\frac{M_\lambda}{\mu}\equiv r,\,m_\chi \equiv \max(\mu,M_\lambda).$
Knowing these corrections allows us to define an effective SUSY scale at leading log precision,
\begin{equation}
m_S^{eff}=\left[m_A^{-\tilde b_\lambda/3} m_{\tilde t}^{8 y_t^4}m_\chi^{4 \tilde b_\lambda/3}\right]^{1/(\tilde b_\lambda + 8 y_t^4)}\,.
\end{equation}
Since $y_t\approx 1/2$ at high scales, the corrections to $m_h/m_W$ are much smaller than in TeV SUSY models. We find
that they can be either positive or negative and are typically below $1$ GeV unless one happens to be in a ``worst-case''
region. This is illustrated in Figure \ref{fig_thresholdhiggsmasses} for both
types of corrections. One sees from the large sensitivity of the new physics scale to $m_t$ that a more precise experimental and theoretical determination of the $\overline{MS}$ top
mass can be crucial for our understanding of UV physics.
\section{UV completions with $\lambda<0$}
A universal feature of the string models we consider here is the 
appearance of an extended SUSY sector at some scale $m_C> m_S$. The 4D 
$D$-Term then becomes one component of an extended scalar potential. As the minimal
example, we consider an $\mathcal N=2$ sector, where the $D$ field is part
of a triplet $\vec P$. The usual MSSM physics is recovered by decoupling two
of these fields in an $\mathcal N=1$ supersymmetric fashion. One 
finds that this decoupling is not exact in the presence of soft SUSY breaking.
In the simplest case, the resulting quartic potential from these effects is given by
\begin{equation}
V_{\Lambda=M} = \kappa^2  \frac{m_s^2}{m_s^2 + M^2} \, 
|H_u H_d|^2 \,.
\end{equation} 
where $M$ is the scale of extended SUSY, and $m_s^2$ the soft breaking parameter.
An interesting consequence is that a negative mass squared parameter will
lead  to a quartic ({\it non-tachyonic!}) instability. One can perform a field
theoretic matching of such an ``unstable'' UV theory to the SM by introducing
a suitable IR cutoff. This is discussed in detail in \cite{Hebecker:2013lha}. This
raises interesting issues for future research. 
It shows that the soft scale can be in the unstable regime and therefore higher than naively expected from
the Higgs mass measurement. Might hierarchies $m_S\ll m_C$
and $m_S\gg$ TeV be connected to vacuum stability at cosmological time scales?
Does the Higgs field still prefer our false weak scale vacuum after inflation in such
scenarios? 

\section*{Acknowledgements}
I would like to thank the organizers of the XXIX-th International Workshop on High Energy Physics in Protvino and the IHEP theory division for their hospitality.

\bibliographystyle{ws-procs975x65}
\bibliography{wsproceedings_knochel}

\end{document}